# Detection and size measurement of individual hemozoin nanocrystals in aquatic environment using a whispering gallery mode resonator

Woosung Kim,[1] Sahin Kaya Ozdemir,[1,‡] Jiangang Zhu,[1] Monifi Faraz,[1] Cevayir Coban,[2] and Lan Yang[1,†]

[1] *Department of Electrical and Systems Engineering, Washington University, St. Louis, MO 63130, USA*
[2] *Laboratory of Malaria Immunology, Immunology Frontier Research Center, Osaka University, Osaka 565-0871, JAPAN*
[‡] ozdemir@ese.wustl.edu  [†] yang@ese.wustl.edu

**Abstract:** We, for the first time, report the detection and the size measurement of single nanoparticles (i.e. polystyrene) in aquatic environment using mode splitting in a whispering gallery mode (WGM) optical resonator, namely a microtoroid resonator. Using this method we achieved detecting and measuring individual synthetic hemozoin nanocrystals―a hemoglobin degradation by-product of malarial parasites―dispersed in a solution or in air. The results of size measurement in solution and in air agree with each other and with those obtained using scanning electron microscope and dynamic light scattering. Moreover, we compare the sensing capabilities of the degenerate (single resonance) and non-degenerate (split mode, doublet) operation regimes of the WGM resonator.

# 1. INTRODUCTION

Rapid detection and measurement of biological and synthetic nanoparticles, bioanalytes, and nanoscale biological material in ultra-small sample size and with single nano-object resolution are of significant importance for diagnostic and prognostic applications, environmental monitoring, food safety, and homeland security. In recent years, different sensor platforms exhibiting enhanced detection capabilities, based on monitoring the shift in the frequency of optical, mechanical and plasmonic resonances, have been developed [1-3]. Here, we report, for the first time, the use of resonance mode splitting in a whispering-gallery-mode (WGM) microtoroidal optical resonator placed in aquatic environment as a platform for label-free detection and measurement of individual nanoscale objects. In particular, we present detection and size measurement of polystyrene (PS) nanoparticles and synthetic hemozoin nanocrystals with single particle/crystal resolution.

Hemozoin is a crystalline metabolite of *Plasmodium* parasites, a causing agent of malaria disease. Despite global eradication strategies all over the world, malaria still kills almost 1 million people each year, urging rapid sensitive diagnosis techniques and effective treatment regimens. It has been recently postulated that detection of hemozoin crystals can be readily exploited for rapid diagnosis of malaria parasites, because hemozoin is continuously produced and released from red blood cells and captured by immune cells of the host during blood circulation of *Plasmodium* parasites [4-6]. Hemozoin is formed inside the acidic digestive vacuoles of intraerythrocytic parasites. The parasites utilize host-hemoglobin and release free-heme ($Fe^{2+}$-protoporphyrin IX) which is subsequently converted into insoluble crystals called hemozoin ($Fe^{3+}$-protoporphyrin IX) as a result of the detoxification process. Recent reports have revealed that optical detection techniques such as third harmonic generation imaging (THG) and tip-enhanced Raman scattering (TERS) can detect hemozoin crystals inside the infected blood cells [7, 8]. However, it is obvious that the level of the signal obtained in these techniques is very much dependent on the number of parasites and hence on the amount of hemozoin, i.e., the larger the amount of hemozoin is, the higher the detected signal intensity is. Moreover, the level of hemozoin circulating in immature forms is very low, making it difficult to detect. Thus, there is a clear need to improve the detection limits of hemozoin crystals to the single nanocrystal/molecule resolution to detect even trace amounts of hemozoin circulating in the blood or serum.

The interests in WGM microresonators have been continuously increasing since the first reports on the detection of streptavidin-to-biotin binding [3] and DNA hybridization [9] by monitoring the resonant frequency shifts in microsphere resonators. These initial works were followed by the demonstrations of single virion and single particle detection using microspheres [10]. The underlying physics of these works rely on the shift of the resonance frequency due to a change in the effective polarizability of the resonator-surrounding compound system upon the entering of a particle, molecule or virion into the resonator mode volume [10, 11]. In spectral shift method (i.e., also known as reactive shift method), individual events are detected accurately; however, size measurement of detected individual particles is hindered by the dependence of the amount of resonance shift on the position of the particle in the mode volume of the resonator. Thus, statistical techniques are employed, e.g., maximum frequency shift observed in an ensemble measurement is used to estimate the size of particles in the ensemble [12]. Statistically significant number of measurements should be performed to have high confidence in assigning an observed shift to be the maximum shift. Moreover, size estimation from resonance shift requires the knowledge of resonator mode volume.

Recently, mode splitting technique has been proposed and used as a highly sensitive and robust platform to detect individual nanoscale materials and take single shot size measurement one-by-one on each detected material. In contrast to the reactive shift method, in mode splitting method, in which a single resonance mode splits into two when a nanoscale object, such as a nanoparticle or a virus, enters the mode volume, size of each detected object can be estimated in a single-shot measurement; neither the position of the object in the mode volume nor the prior information of the WGM volume is required for the measurement [13]. The early works performed in dry environment were followed by the demonstration of mode splitting in microtoroid [14] and microsphere [15] resonators in aquatic environment. However, whether the splitting was induced by a single particle or not could not be confirmed. Recently, Lu *et al*. have shown single particle induced mode splitting in a microtoroid placed in water [16]. In order to minimize the noises introduced into the transmission spectra by spectral fluctuations of the tunable laser, whose wavelength is scanned to monitor the resonance shifts and mode splitting, the authors used a thermally stabilized reference interferometer. He *et al.*, on the other hand, reported single particle induced mode splitting in water using an active microresonator driven above lasing threshold [17]. However, none of these works provided the polarizability and hence the size measurement of each detected nanoparticle in aquatic environments.

Here, for the first time, we report the detection and size measurement of individual nanoparticles (PS and synthetic hemozoin) in aquatic environment using mode splitting in a passive (no active gain medium) microtoroid resonator without the need for stabilization or noise cancellation interferometers. Moreover, we quantify the differences between the spectral shift and mode splitting methods and the regimes of quality ($Q$) factor and particle size where one or both of these methods can be effectively used.

## 2. MATERIAL AND METHODS

### 2.1. Materials

Polystyrene nanoparticles were purchased from Thermo Scientific with mean radius of *R=75nm* with standard deviation of *σ=2.2nm*. The coefficient of variation (CVs) defined as the ratio of the standard deviation to mean value, for these particles is *2.9%*. Ultrapure DI water was obtained from Sigma-Aldrich. Silica wafers used for the fabrication of the microtoroids were purchased from WRS Materials.

### 2.2. Methods

#### 2.2.1. Microtoroid and taper fiber fabrication process

Microtoroids were fabricated from silica-on-silicon wafer through a three step process as described previously [18]; (i) Standard photolithography for patterning circular silica disks on silicon wafer, (ii) Xenon difluoride (XeF) for selective isotropic etching of silicon to form undercut structure with silica disk over silicon pillar, and (iii) $CO_2$ reflow to transform the disk to a torus forming silica microtoroid on silicon pillar. A microtoroid is defined by its major and minor diameters. For our experiments, we fabricated microtoroids of minor diameters *d=7~10μm* and major diameters *D=50~80μm*.

#### 2.2.2. Taper fiber fabrication process

Taper fiber for coupling light into and out of the microtoroid resonator was fabricated by heating-and-pulling a single mode silica fiber under hydrogen flame. The jackets of single mode silica optical fibers were removed and the exposed clad was cleaned with isopropyl alcohol. The fiber was then placed on a home-made holder with fiber clips at two ends, and positioned above hydrogen flame. Pulling speed and distance were controlled by two servo motor controllers. Tapering process was continuously monitored under microscope. Optical fiber taper with diameters approximately *1μm* was fabricated.

#### 2.2.3. Preparation of synthetic hemozoin

Synthetic hemozoin, chemically shown to be identical to natural hemozoin [19], was purified from hemin chloride using an acid-catalyzed method which produces smaller and *~200nm* in size homogenous crystals as described earlier [6, 20] Briefly, *45mg* hemin chloride was dissolved in *4.5ml 1N* NaOH and neutralized with *1N* HCl. With the addition of acetic acid (until the pH reached *4.8*, at a constant temperature of *60ºC* with magnetic stirring), the mixture then was allowed to precipitate at room temperature overnight. The precipitate was subjected to a wash with *2%* SDS-buffered with *0.1M* sodium bicarbonate (pH *9.1*) and subsequent extensive washes with *2%* SDS, and then six to eight washes with distilled water. A stock solution in distilled water was prepared and stored at *4°C*. The hemozoin was dried, weighed and the concentration was calculated in either *mM* or *mg/ml*. FTIR and powder X-ray diffraction pattern of the synthetic hemozoin confirmed the chemical composition and structure as previously described [20]. Figure 1 shows the different stages after the infection of an erythrocyte with a rodent parasite *P. yoelii* and the FESEM images of synthetic hemozoin visualized using ultra-high-resolution FESEM.

*2.2.4. Delivery of the particles and hemozoin crystals to the resonator mode volume*

For experiments performed in water, the solution containing the hemozoin crystals or the nanoparticles was continuously pumped into and out of the aquatic chamber (see Fig. 2) used in the experiments at a rate of *1ml/min*. We chose a flow path such that the particles pass proximity of the resonator, entering its mode volume or occasionally adsorbed to the resonator surface.

For hemozoin and PS particle experiments performed in air, we used the setup reported in our previous work [13], which consists of an atomizer and a differential mobility analyzer (DMA) connected to a micro-nozzle (diameter: *20μm*). The DMA is one of the most commonly used equipment that classifies and measures the size of nanoparticles, based on their electrical mobility [13, 21]. The atomizer is used to disperse the nanoparticles or the hemozoin crystals from the particle suspension. Droplets produced by the atomizer are passed through a $Po^{210}$ radioactive neutralizer which removes the electrical charges on the particles and then sent to a diffusion dryer with silicone-gel as the desiccant for removing the solvent in the droplets. The particle stream is then sent through a $Kr^{85}$ radioactive particle charger to place a well-defined charge distribution on the particles as a function of their size. Afterwards, the particles flow to the DMA which consists of a flow channel between an assembly of two concentric cylindrical electrodes, separated with an air gap. An electric field is applied between the electrodes. The particle flow, introduced near one of those electrodes, is joined by a larger sheath flow. As the particles are carried along the channel by this combined flow, the particles of appropriate charge polarity migrates across the channel under the voltage difference between the electrodes. The rate of the migration depends on the electrical mobility of the particles, which in turn, depends on the size and electrical charge of the particle. If all the particles in the flow have the same charge, then particles having a specific mobility will be all of the same size. Moreover, the smaller the particle and/or the higher the electrical charge is, the higher the electrical mobility is. A small slit then allows the particles having a specific electrical mobility within a narrow range to exit. Thus, particles whose electrical mobility matches the position of the outlet slit at a certain applied voltage pass the DMA and classified. For a given DMA configuration, the electrical mobility in the DMA to outlet the particles at a certain location is given as

$$Z_p = \frac{q_c + q_m}{4\pi \Lambda V} \qquad (1)$$

where $q_c$ and $q_m$ are the flow of sheath air at the DMA entry and the exit, respectively, and $\Lambda$ is a constant depending on the DMA configuration given as

$$\Lambda = \frac{L}{\ln(r_2 / r_1)} \qquad (2)$$

with *L*, $r_1$, and $r_2$ denoting the length of the DMA, and the radii of the inner and outer electrodes from the DMA capacitor, respectively. On the other hand, the electrical mobility, $Z_p$, of a singly charged particle with diameter *d* is given as

$$Z_p = \frac{e C_d}{3\pi \eta d} \qquad (3)$$

where *e* is the charge of an electron, $\eta$ is the viscosity of the gas and $C_d$ is the Cunningham slip correction which approaches to one for a particle larger than the mean free path of the gas and increases with decreasing particle size. For particles exiting the DMA, Eqs. 1 and 3 are equal. Thus by changing the applied voltage, *V*, particles within a certain diameter range can be made to exit the DMA.

In our experiments, the particles exiting the DMA are then deposited through the micro-nozzle onto the microtoroid resonator which was placed a few hundreds of microns away from the nozzle. The flow rate was set to be *0.02cm$^3$/s* so that there is no more than one particle deposition event within the detection window of *0.1s*. Moreover the concentration of the hemozoin crystals or the nanoparticles in the solution was adjusted to be *~100fM* to further suppress simultaneous deposition of multiple particles. With these settings, we observed on average one particle or crystal binding event in every *5s*.

For accurate size measurements, it is important that no aggregation takes place, because aggregations will lead to larger size readings. In order to prevent or minimize aggregations, we used water bath sonication for 5 minutes prior to each experiment performed in water and in air. Moreover, since we have worked with samples of very low concentration (*~100fM*), the likelihood of collisions and interactions which may lead to aggregation is low (i.e., the higher concentration of the particles or crystals in the suspension, the higher the probability of aggregation). Indeed, we also performed dynamic light scattering (DLS) and SEM measurements and could not see strong indication of aggregation (i.e., DLS size measurement distribution indicates the degree of aggregation).

*2.2.5. Experimental setup*

The experimental setup used during the measurements in aquatic environment is similar to our previous work [14] as shown in Fig. 2. The chip containing the microtoroid resonators is immersed in an aquarium which is placed on 3D nanocube to finely tune the distance between the resonator and the taper fiber. Light from a *670nm* tunable diode laser was delivered to the resonator through a fiber taper. The output wavelength of the laser is continuously scanned at a rate of *5Hz*. Among the four tunable lasers (*660nm*, *980nm*, *1460nm* and *1550nm* wavelengths) we chose the visible wavelength laser (*λ=670nm*) because light absorption in water is weaker in the visible band than the near-infrared band. A silicon PIN photodetector with *125MHz* bandwidth connected to an oscilloscope was placed at the output end of the optical fiber to monitor the transmission spectra. When the laser wavelength satisfies the resonant condition, the light is coupled into the WGM of the microtoroid resonator causing a resonance dip in the transmission. The wavelength at which the dip occurs corresponds to the resonance wavelength. The measurement system was controlled by a computer using Labview. When a nano-scale object enters the field of the resonator, the resonance splits into two forming a doublet in the transmission spectra. A two-Lorentzian curve fitting is applied as soon as the transmission is acquired to the computer to obtain the resonance frequencies and linewidths of the doublets. After obtaining these parameters, the polarizability and then the size of each detected nano-object are estimated.

*2.2.6. Evaluating the performance of mode splitting method*

The performance of mode splitting method in estimating the size of each detected particle is evaluated by comparing the estimated sizes with the size distribution obtained from dynamic light scattering (DLS, Malvern Zetasizer Nano) system and from scanning electron microscope (SEM, Nova 2300, FEI) images, as well as with the data provided by the manufacturers of the nanoparticles. Moreover, for the measurements performed in air, we used a DMA which selects the deposited particles based on their electrical mobility with approximately *5%* size deviation. In the following, we present the basics of size measurement by DMA, DLS, and SEM.

***Differential Mobility Analyzer (DMA)*** selects particles of certain electrical mobility. It can be also used to make size measurements. From Eqs. 1 and 3, we have

$$d = \frac{eC_d}{3\pi\eta} \frac{4\pi\Lambda V}{(q_c + q_m)} \quad (4)$$

for a given DMA configuration and applied voltage. Since $C_d$ depends on the size of the particles, this relation should be iteratively solved to estimate the size. The DMA used in our experiments selectively delivers nanoparticles within approximately *5%* size deviation based upon the electric mobility of the nanoparticle [22]. We should emphasize here once more that that the size selected and/or estimated in a DMA is the electrical mobility size.

***Dynamic Light Scattering (DLS)*** is a well-established technique for measuring the size and size distribution of small, typically sub-micron, particles (e.g., dielectric or metallic nanoparticles, molecules, proteins, polymers, etc.) dispersed in a liquid environment [23]. Particles in a dispersion are in a constant, random motion known as Brownian motion (i.e., the larger the particle, the slower the Brownian motion). When such dispersion is illuminated by a coherent light source (i.e., a laser), time-varying intensity fluctuations, 'dynamic speckles', in which the position of each speckle is seen to be in motion, are observed due to the destructive or constructive interference between the light scattered from moving particles whose distances from each other are constantly varying. The rate at which these fluctuations take place yields the velocity of the Brownian motion and hence the particle size. Thus, in principle, DLS measures the Brownian motion, which is then related to the size through the Stokes-Einstein relationship. In a typical DLS measurement, the rate of changes in the scattered light intensity can be obtained by comparing the intensity $I(t)$ at time $t$ to the intensity $I(t+\tau)$ at a later time $t+\tau$. Making this comparison over a long measurement time $T$ and for different $\tau$ yields the intensity correlation function $G_2(\tau)$

$$G_2(\tau) = \frac{1}{T}\int_0^T I(t)I(t+\tau)d\tau \quad (5)$$

which describes the dynamics of the particle motion but does not yield information on their correlated motions. This information can be obtained from the electric field correlation function $G_1(\tau)$

$$G_1(\tau) = \frac{1}{T}\int_0^T E(t)E(t+\tau)d\tau \quad (6)$$

which is related to the measured $G_2(\tau)$ through Seigert relationship

$$G_2(\tau) = B\left[1 + \beta|G_1(\tau)|^2\right] \quad (7)$$

and has an exponential form with decay rate $\Gamma$

$$G_1(\tau) = e^{-\Gamma\tau} \quad (8)$$

for a system undergoing Brownian motion. In Eq. 7, *B* and *β* are baseline and instrumental response which are determined experimentally. The decay constant $\Gamma$ is related to diffusivity (*D*)

$$\Gamma = Dq^2 \quad (9)$$

where

$$q = \frac{4\pi n}{\lambda} \sin\frac{\theta}{2} \qquad (10)$$

reflects the distance the particle travels. Thus, *D* can be determined from experimentally measured *Γ* if the angle *θ* and the refractive index *n* are known. Finally, the particle size for spherical particles is extracted from the Stokes-Einstein relation, which relates *D* to thermodynamics and hydrodynamics as

$$D = \frac{kT}{6\pi\eta r} \qquad (11)$$

where *k* is the Boltzman's constant, *T* is the temperature, *η* is the viscosity and *r* is the radius of the particle. Thus, size is determined from *D* if *T* and *μ* are known. If the system is monodisperse, then the mean effective diameter of the particles is determined. It should be noted here that size estimated in DLS is based on the translational diffusion of the particle within the dispersion, thus it is referred to as the hydrodynamic size (i.e., the size of a sphere having the same diffusion coefficient as the particle being measured).

***For SEM measurements***, a number of particles are deposited onto a microtoroid resonator using the DMA and then the resonator was taken to SEM. Particle size (geometric size) was estimated by processing the acquired particle images with a built-in toolkit of the SEM. This is done by calculating the circle equivalent diameter by measuring the area of the image of the particle and then back-calculating the diameter of a circle with the same area. This method depends on the edge-contrast settings and on which view is captured. Thus, obtained results may not be directly comparable with other size measurement methods, especially when the particles are non-spherical.

It is clear that each of the above methods gives the size of a particle just by one quantity (e.g., a number and a unit) regardless of its shape. If all the particles measured have the same shape, this is unambiguous. However, when the particles have different and/or irregular shapes, one single quantity to describe the size may cause ambiguity. However, usually the shapes of the particles to be measured are not given or are not known exactly. Thus, in particle size measurement and aerosol technology, an equivalent diameter or radius has been introduced and used to give the particle size by one quantity regardless of their different shapes [24]. Using equivalent diameters, various irregularly shaped particles can be evaluated on the basis of a single consistent measurement. An *equivalent spherical diameter (or radius)* is the diameter (or radius) of a sphere which has a certain size-dependent property with the particle being measured. This size-dependent property depends on the measurement technique used for the determination of the particle size. Different techniques measure different properties of the particle, and thus produce different size distributions for identical samples. In DLS the estimated hydrodynamic size for non-spherical particles is the size of spherical particles which has the same translational diffusion as the particle being measured. Thus, changes in the particle shape, such as the length of a rod-like particle, affect the diffusion speed of the particle. On the other hand, different shape changes, such as the diameter of a rod-like particle, which hardly affects the diffusion speed, may not be detected and hence the size estimation will not change. In DMA, the estimated size is the equivalent electrical mobility diameter that is the size of spherical particles which has the same electrical mobility as the particle under test. In SEM or TEM, the estimated size can be the equivalent circle diameter, which is the diameter of a sphere whose projected area is the same as the projected particle image.

Thus, in many applications, where the exact shape of the particles are not known, the estimated size is the size of a sphere which has the same measured property (e.g., translational speed, polarizability, view or image) as the particle under test. Thus, in this work we follow the same methodology and assign the measured size as the size of a spherical particle having the same measured polarizability as the particle being measured using mode splitting in a WGM microtoroid resonator. In short, here we measure the equivalent polarizability diameter.

## 3. RESULTS AND DISCUSSIONS

### 3.1. Basis for WGM optical resonator based sensors

A WGM resonator supports the propagation of light via total internal reflection if the optical path length is an integer multiple of the wavelength of the WGM light, that is $2\pi n_{eff} R = m\lambda$ where $n_{eff}$ is the effective refractive index of the medium experienced by the WGM field, $R$ is the radius of the resonator, $m$ is an integer number, and $\lambda$ is the wavelength of the light. The evanescent tail of the WGM field probes the surroundings of the resonator and responds to the changes in the surroundings. The interaction of the evanescent WGM field and the analyte, which can be a biomolecule, a synthetic or biological particle, an aerosol, a protein, etc., enables the detection of the analyte or the changes taking place in the analyte without labeling.

Any changes in the surroundings, due to either a change in the refractive index or analyte molecules, particles or proteins entering the WGM evanescent field, lead to excess polarizability which is reflected in the transmission spectrum of the resonator either as a resonance shift (reactive shift) generally accompanied by linewidth broadening or as mode splitting. Figure 3 shows two typical experimental spectra depicting spectral shift (Fig. 3(a)) and mode splitting (Fig. 3(b)) phenomena observed with a microtoroid resonator.

WGM resonators, in particular the microtoroid resonators used in our experiments, have very high $Q$ factors ($>10^7$), thus photons circulate over thousands of times within the circular boundary of the microtoroids, immensely increasing the interaction length between the evanescent field and the analyte. At the same time, the microscale mode volumes and very small mode areas of microtoroid resonators lead to unprecedented levels of light intensity, which further enhances the interaction, and thus improves sensitivity.

### 3.2. Theoretical basis of detection and measurement of nano-scale objects using mode splitting

WGM resonators support two modes of the same frequency but different propagation directions, i.e., frequency degenerate modes. In the presence of a scattering center (e.g., scatterer, nanoparticle, biomolecule, virus, or any structural inhomogeneity and surface roughness, etc.), the WGM field interacts with the scattering center leading to elastic scattering, which mixes light fields of the counter-propagating modes within the cavity. Meanwhile, a portion of the light is lost to the environment. If the interaction strength is strong enough, the frequency degeneracy of the WGM resonator is lifted [25-27], leading to two spectrally different resonances (Fig. 3(b)). Central frequencies and linewidths of these split resonances are different but are related to each other via

$$2|g| = \frac{\alpha f^2(r)\omega}{V} \qquad (12)$$

and

$$2\Gamma = \frac{2\alpha |g| \omega^3}{3\pi v^3} \qquad (13)$$

which define the spectral distance between the split modes (i.e., interaction strength), and the linewidth difference (i.e., due to scatterer-induced additional loss) between the split resonances, respectively. In these expressions, the effect of the particle is denoted by the particle polarizability $\alpha$ defined as

$$\alpha = 4\pi R^3 n_e^2 \frac{n_p^2 - n_e^2}{n_p^2 + 2n_e^2} \qquad (14)$$

for a spherical particle of radius $R$ and refractive index $n_p$ in a surrounding environment of refractive index $n_e$. The resonator related parameters are the pre-splitting initial angular frequency $\omega$, the quality factor $Q$, mode volume $V$, the speed of light $v$ in the medium, and the WGM field distribution characterized by $f(r)$. It is clear that mode splitting spectra carry the polarizability information from which size can be estimated accurately if $n_p$ is known.

In mode splitting method, there is a transition from single resonance to a doublet (split resonances) or a change in the mode splitting spectra characterized by $2|g|$ and $\Gamma$, if a particle enters the mode volume. Polarizability of the particle is estimated by comparing the spectral properties of the splitting spectra just before and after the arrival of the particle using

$$\alpha = \frac{3}{8\pi^2} \left(\frac{\lambda}{n_e}\right)^3 \frac{|\gamma_N - \gamma_{N-1}|}{|\delta_N - \delta_{N-1}|} \qquad (15)$$

where $\gamma_i$ and $\delta_i$ are the sum of the linewidths and resonance frequencies of the split modes after the arrival of the $i$-th particle, respectively. Zhu *et al*. have used this method to estimate the polarizability or size of gold, polystyrene, and influenza virus in dry environment [28, 29].

The most interesting feature of the mode splitting method is that once the particle is detected, its polarizability can be estimated without knowing where the particle is located in the mode volume. Moreover, since both of the split modes reside in the same resonator, they are affected similarly by the noise sources. Thus, the effects of these noises are automatically minimized, if not eliminated, without need for external referencing or stabilization schemes that increase complexity of the measurements, because one mode acts as a reference to the other mode. Mode splitting then provides a self-referencing platform.

In summary, mode splitting is a result of the scatterer-induced lifting of the inherit frequency degeneracy of WGM resonators. If mode splitting does not take place, then the frequency degeneracy of the WGM survives and transmission spectra show single resonance. Therefore, mode splitting method which relies on the spectral properties of split modes corresponds to "non-degenerate" sensing scheme whereas spectral shift method (reactive shift method) which relies on the shift in resonance frequency of a single mode (i.e., no mode splitting) corresponds to "degenerate" sensing scheme.

### 3.3. What determines the splitting spectra and the sensing performance?

Recent experiments and our study on nanoparticle detection using WGM resonators in dry and aquatic environments have shown that in some experiments only a spectral shift accompanied with linewidth broadening is noticed, whereas in some other experiments clear split modes are observed for the particles of the same size and refractive index [14, 15, 28-30]. This, naturally, brings the question of under what conditions mode splitting or the spectral shift takes place and what is achievable in each of these methods, in other words, what makes one preferable over the other.

The strength of the interaction of the WGM field with the nanoparticle (i.e., whether the particle is located at a strong or weak field region), the amount of excess polarizability (i.e., size of the particle and the refractive index contrast between the particle and the environment), quality factor of the resonator (i.e., linewidth of the resonance) and the noise level (e.g., laser frequency and amplitude fluctuations, wavelength scanning noise, electronic noise, etc) in the system determine whether resonance shift or mode splitting is observed and resolved in the transmission spectrum.

In order to observe mode splitting, the interaction between the WGM field and the scatterer, such as a nano-scale object or material, should be strong enough to overcome the total loss (e.g., sum of the resonator and particle related losses) of the system. This condition can be re-stated as $2|g|>\Gamma+\omega/Q$, from which mode splitting quality factor is defined as $Q_{sp}=|2g|/(\Gamma+\omega/Q)$ [31]. In practical settings, various noise sources, such as laser intensity and frequency fluctuations, detector noise, taper-cavity gap fluctuations, and signal processing noise strongly affect the accuracy of measuring the resonance frequencies and the linewidths of the split modes. This, in turn, affects the $Q_{sp}$ and the resolvability of mode splitting, as well as the accuracy of single-shot size measurement using mode splitting. In the presence of noise, higher $Q_{sp}$ values are required to resolve splitting and to make highly accurate size measurements.

Figure 4(a) shows the expected the transmission spectra for $Q_{sp}$ values in the range *0.6−1.2*. It is seen that as $Q_{sp}$ increases, the features of split modes become clearer. Thus, both the linewidths and the resonance frequencies of the split modes can be extracted using the curve fitting algorithm with high accuracy. For smaller values of $Q_{sp}$, it becomes difficult to extract the features accurately as the error in Lorentzian fitting to extract the resonance frequencies and the linewidth broadening increases with decreasing $Q_{sp}$. This leads to error in size estimation. For example, when $Q_{sp}=0.6$, only one resonance mode with strongly broadened linewidth is observed. In such a case, either mode splitting is buried within the broad linewidth or it really has not taken place. We cannot discriminate between these two cases; therefore, it is difficult to arrive at an accurate conclusion.

We measured noise level in our system in aquatic environment before splitting took place (i.e., before nanoparticle binding while there was still a single resonance in the transmission spectra) in order to quantify its effects on the measurement results. We continuously acquired transmission spectra and performed Lorentzian curve fitting on selected resonance modes, and analyzed the statistics of the measured linewidth and resonance frequency. Resonance frequencies were repeatedly extracted with high accuracy for the selected modes. However, the same cannot be said for the linewidth measurements. We observed that the main source of error in our scheme is the linewidth measurement.

Standard deviations $\sigma$ of linewidth measurements for resonance modes when the microtoroid was immersed in aquatic environment are depicted in Fig. 4(b) and Fig. 4(c). It is seen that for low $Q$ modes (i.e., modes with larger linewidths), $\sigma$ is larger and the dependency can be approximated by *$\sigma=0.47\times\mu^{0.62}$* where $\mu$ is mean value of the measured single mode linewidth. Figure 4(d) depicts the dependence of the coefficient of variation, which is defined as the ratio of standard deviation $\sigma$ to the mean value $\mu$, on the $Q$ value. The best fit to this experimentally obtained data gives the relation *$\sigma/\mu=3\times Q^{0.34}\times 10^{-4}$*.

For the range of nanoparticle sizes we aim to detect, linewidth broadening is much smaller than the amount of mode splitting, that is $|2g|>\Gamma$, and the quality factor of the resonance modes does not change much. Considering this point together with the results of the above experiments, we conclude that measuring $2g$ or the spectral shift is easier than measuring $\Gamma$ in our system. Therefore, in the presence of noise linewidth measurement is the limiting factor for the accuracy of particle size measurement.

In order to accurately estimate the size of a detected nanoparticle, linewidth broadening $2\Gamma$ induced by the particle should be greater than the noise level (standard deviation) of measuring the linewidth difference. If the noise sources are neglected, we can measure any *$\Gamma>0$* without any problem. In the case of noise, however, the mode splitting observability criterion becomes much stricter, because in this case the scatterer induced linewidth broadening should be larger than the fluctuations of the linewidth differences of the split modes.

Assuming that the coefficient of variation of linewidth measurements of each of the split modes is $\sigma/\mu$, we can set a loose bound on $\Gamma$ as *$\Gamma>(\omega/Q)(\sigma/\mu)$*. Consequently, we find that $Q>(\omega/\Gamma)(\sigma/\mu)$ should be satisfied to accurately measure the linewidth broadening (i.e., accurately estimating the polarizability and size of the particle) in a noisy system. Using the above relation, we find that for accurate estimation of linewidth broadening, quality factor of the resonance mode before the binding of the particle should satisfy $Q>(3\times10^{-4}\times\omega/\Gamma)^{1.51}$.

It is now clear that accurate size measurement in aquatic environment using split modes requires two conditions to be satisfied:

(i) $Q_{sp} \geq \eta$ where $\eta$ is a real number below which mode splitting cannot be resolved and single resonance appears in the transmission spectra. The value of $\eta$ depends on the experimental conditions regarding the particle, the resonator and the noise in the system. $\eta \geq 1$ implies that there is more noise to deal with to clearly resolve split modes.

(ii) $Q>(\omega/\Gamma)(\sigma/\mu)$ to assure accurate estimation of linewidth broadening and hence the polarizability and/or the size of the particle. Using (i) and (ii), we determine the smallest nanoparticle size $R$ that can be detected and accurately measured using a WGM resonator of quality factor $Q$. Moreover, we can determine the parameter space bounded by $Q$ and $R$ in which the WGM resonator works in single or split mode regimes, and in which accurate size estimation is possible with mode splitting. There are four possible cases for the interaction between WGM resonators and nanoparticles:

*(a) Both (i) and (ii) are satisfied simultaneously.* The sensing platform operates in the split mode regime with highly accurate size measurement.

*(b) Only (i) is satisfied.* Presence of particle is detected by splitting, however, size measurement will be erroneous due to the noise in linewidth measurement.

*(c) Only (ii) is satisfied.* There is no splitting or splitting cannot be resolved as it is hidden in the broadened linewidth. Sensor operates in single mode regime and size measurement is possible if the shift in the single resonance can be detected. Note that none of the previously reported works based on reactive shift method has considered looking at the change in the linewidth of the single resonance mode. Instead, only shift in resonance has been considered. If the change in linewidth is measured in addition to the frequency shift, size measurement can be performed in such a way similar to mode splitting method.

*(d) Neither (i) nor (ii) is satisfied.* Sensor is in single mode regime (either there is no mode splitting or splitting cannot be resolved) and linewidth measurement is erroneous. Particle might be detected via the changes in resonance frequency or linewidths, but accurate size measurement is not possible.

Since the size of the WGM microresonator affects the noise level and $Q_{sp}$, the boundaries of the four regimes discussed above should be different for resonators of different sizes. Figure 5 depicts the boundaries between different regimes for two microtoroids of different sizes assuming that the particle is at field maximum, and taking into account the observed noise level in our system when the measurements were done in aquatic environment. It is clear that single mode operation takes place for low $Q$ resonators and nano-scale objects with smaller $R$. For a fixed particle size, mode splitting requires larger $Q$. In order to push the limits of measurable particle size to smaller values, resonators with higher $Q$ values are required, which, on the other hand, naturally leads to mode splitting. For example, from Fig. 5(a) we see that with a resonator of $Q \geq 10^7$, mode splitting with $Q_{sp} \geq 1$ can be obtained for PS particles $R \geq 35nm$, implying that we can detect these particles but size measurement may contain error. For the present noise level we have in our system $Q>(\omega/\Gamma)(\sigma/\mu)$ is satisfied only for PS particles $R \geq 50nm$, implying that only these particles can be detected and measured with high accuracy. The amount of error decreases as $Q_{sp}$ and $Q$ becomes larger.

The four different regimes (a)-(d) characterized by the conditions (i) and (ii) are plotted in Fig. 5 for the experimentally observed noise level, and for $\eta=1$ and $\eta=0.5$. For $\eta=1$ in Fig. 5(a), the operating regimes (a) and (b) are labeled as (1) and (6), respectively, whereas (c) corresponds to the union of regions labeled as 2 and 3, and (d) is the region formed by the union of (4) and (5). For $\eta=0.5$, on the other hand, (c) and (d) corresponds to regions (3) and (4), respectively, whereas (a) and (b) fall within the regions represented by the union of (1) and (2), and that of (5) and (6), respectively.

Figure 5 clearly shows the dependence of the detection capability and the size measurement accuracy of mode splitting method on the resonator $Q$, PS particle size $R$, noise level in the system, and resonator size (i.e., mode volume). Decreasing the noise level in our system will improve our detection and size measurement limit. Also comparison of Figs. 5(a) and 5(b) reveals that a smaller resonator leads to better detection limit, e.g., for $Q=10^6$ and $\eta\geq1$, lower detection limit for the toroid $D=80\mu m$ is $R=77nm$ (Fig. 5(a)) whereas for the toroid $D=53\mu m$ it is $R=65nm$ (Fig. 5(b)).

In Fig. 5(a), we see that linewidth measurement noise is a limiting factor for accurately measurable size. For resonance shift method this limit is around $R=64nm$. To measure smaller particles, higher $Q$ resonators should be used. In this case, one will inevitably enter the mode splitting region. For example, to accurately measure a particle of $R=50nm$, one cannot use resonance shift method. However, if $Q$ is increased to around $10^7$, mode splitting method will allow accurate measurement. Detection limit, on the other hand, is determined by the noise of resonance frequency measurement, which is much lower than the noise in linewidth measurement. For the detection limit, we have two cases. First, if initially there is no mode splitting, resonance shift method, in principle, has lower detection limit, because for the same resonator, resolvability condition of mode splitting requires higher quality factor. Second, if there is an initial splitting (e.g., intrinsic splitting or splitting induced by a previous particle), mode splitting is more sensitive because both modes of the splitting will be affected by an incoming particle, experiencing a total shift of twice that of the shift of a single mode case. Therefore, signal arising from the same particle is larger in mode splitting case, which implies that one can lower detection limit beyond that of resonance shift.

### 3.4. Detection and size measurement of individual PS nanoparticles in aquatic environment using mode splitting

Here, we report the first experimental demonstration of mode splitting based nanoparticle detection and size measurement in aquatic environment. Previously, we have reported mode splitting in aquatic environment; however, we could not reach single nanoparticle resolution, thus size measurement could not be performed. On the other hand, Ref. [17] has reported single nanoparticle resolution without being able to perform single-shot size measurement of the binding particles. In this work, we have successfully detected and measured the size of individual PS nanoparticles using mode splitting in aquatic environment.

Figure 6 shows the results of experiments performed in aquatic environment for PS nanoparticles of $R=75nm$ using a microtoroid resonator of major diameter $D=80\mu m$. The resonance mode used in the experiments was initially degenerate (no observable mode splitting) and had quality factor $Q=6\times10^6$. Experiments were performed using a solution of PS nanoparticles in deionized water with a concentration of approximately $1nM$. The solution was injected into the chamber continuously while the transmission spectra were monitored and size was estimated continuously. Figure 6(a) shows a series of particle binding events reflected in the splitting amount $2g$ as discrete upward and downward jumps above the noise level. The inset depicts two typical largely separated and well-resolved split resonances in the transmission spectra. Splitting quality factors of the events in Fig. 6(a) lie in the range $0.02\leq Q_{sp}\leq1.42$. We find that the estimated size of each detected particle differs although the average value of $<R>=75.8nm$ is in good agreement with the size distribution provided by the manufacturer as $R=75\pm2.2nm$. We find that the results in Fig. 6(b) obtained from single-shot size measurement of each detected nanoparticle in the aquatic environment have a good match with the results of DLS measurements shown in Fig. 6(c), and those of mode splitting experiments in air which yielded $<R>=76.5nm$ with a standard deviation of $9.2nm$. We tested particles from the same batch under SEM and estimated an average size of $75.1nm$. Estimated average sizes are in good agreement in all the methods tested in this study.

In Fig. 6(b), there are three cases in which the estimated radii $R=188.8nm$, $R=203.8nm$ and $R=317.9nm$ are far larger than the nominal size of $R=75nm$ for this batch of particles. When we looked closely to the mode splitting spectra for these three events, we found the corresponding $Q_{sp}$ values as $Q_{sp}=0.49$, $Q_{sp}=0.31$, and $Q_{sp}=0.70$. This suggests that although for the events with $Q_{sp}=0.49$ and $Q_{sp}=0.70$, the condition (i) is satisfied, but the condition (ii) is not, whereas for the event with $Q_{sp}=0.31$, neither (i) nor (ii) is satisfied. Thus, for these events size estimation errors are expected. Another reason for the estimation error may be the aggregation of particles forming larger particles.

Results presented in Fig. 6 and the discussions above reveal that the conditions given in (i) and (ii) are very important for accurate size measurement. We looked closer at Fig. 6(b) and quantified how the choice of $\eta$ in condition (i) affects the size measurement in order to set a threshold $\eta$ above which size measurement will be highly accurate, and below which size measurement is unreliable. We found that by taking only the events for which $Q_{sp} \geq 0.43$, size measurement yields *73.3nm≤R≤75.9nm*. In the distribution depicted in Fig. 6(b), we observed that the events falling within *1σ* of the mean *<R>=75.8nm* have *0.26≤$Q_{sp}$≤1.42* leading to estimated sizes in the range *40.4nm≤R≤102.3nm,* whereas for those within *2σ*, lower bound of $Q_{sp}$ decreases down to *0.20* shifting the estimated sizes to the range *19.5nm≤R≤143.4nm*. For $Q_{sp} \leq 0.43$, estimation error increases.

### 3.5. Detection and measurement of hemozoin in air and in aquatic environment

We report results of a series of experiments done to test the performance of mode splitting method in detecting and estimating the size of hemozoin crystals both in air and in aquatic environment. This is the first time hemozoin is detected and measured using a WGM optical microcavity. In mode splitting and DLS experiments, we take the refractive index of hemozoin as *1.44* as reported in the literature [32].

Our first set of experiments with hemozoin crystals were performed in dry environment. First, we deposited hemozoin particles on a microtoroid resonator using the DMA, and analyzed it under SEM. Figure 7(a) shows an SEM image of the microtoroid with synthetic hemozoin particles. The estimated sizes of the two hemozoin crystals in the image are *R=143.2nm* and *R=141.0nm*. Next, we continuously deposited hemozoin crystals onto a microtoroid resonator one-by-one and monitored the transmission spectra. As expected with the first particle, mode splitting took place and with each subsequently detected particle mode splitting spectra were modified. In Fig. 7(b), we depict nine consecutive single hemozoin binding events where discrete upward and downward jumps are clearly observed. Each of these discrete jumps corresponds to one hemozoin binding event. We measured the size of each detected hemozoin crystal and estimated the mean size, *R=159.2nm*. Note that the size obtained from the mode splitting method here is the size of a spherical particle having the same refractive index as hemozoin and leading to the same excess polarizability as the deposited hemozoin crystal.

The second set of experiments was performed in aquatic environment. Using samples from a hemozoin solution of concentration *~1mM* stirred by a sonicator, we first performed DLS measurements whose results are shown in Fig. 7(c). The mean size of hemozoin crystals in this solution was calculated as *R=160.7nm* using the mean values obtained from multiple runs of DLS. Finally, we performed mode splitting experiment in aquatic environment using hemozoin crystals. The solution used in this experiment has a concentration of *~100fM*. Figure 7(d) depicts nine consecutive discrete upward or downward jumps in *2g* corresponding to nine hemozoin binding events. We estimated the average size of hemozoin crystals as *R=160.9nm*.

The underlying physics of our experiments is the induced polarization by the cavity field on the nanoparticle or the hemozoin crystal staying within the mode volume. The strength of the induced polarization is a function of the properties of the particle (e.g., shape, dimensions, orientation, and refractive index) and the cavity field (e.g., polarization: transverse electric (TE) or transverse magnetic (TM)) interacting with the particle. In general, it is expected that TE and TM waves will induce different polarizabilities, because they have different field components: TE wave has only a tangential component whereas TM wave has both tangential and normal components. For microtoridal resonators with small minor diameters, non-transverse electric field components are relatively small when compared to the transverse component and hence can be neglected. Thus, for such resonators polarizing fields are tangential and normal components of TE and TM modes. Similarly, the orientation of the particle or the crystal on the resonator surface has tangential and normal components; therefore, it is expected that for particles of irregular shapes, different orientations will result in different polarizabilities for a fixed field mode.

For a particle deposited on the resonator with a fixed orientation, we can define polarizability to have a tangential $\alpha_t$ and a normal $\alpha_n$ component such that TE and TM modes, respectively, induce $\alpha_t$ and $\alpha_n$. Note that polarizability of a spherical particle is the same for TE and TM modes and for all of its possible orientations, provided that the microsphere is completely within the cavity field, and it is simply given as

$$\alpha = 4\pi R^3 n_e^2 \frac{n_p^2 - n_e^2}{n_p^2 + 2n_e^2} = 3V_p n_e^2 \frac{n_p^2 - n_e^2}{n_p^2 + 2n_e^2} \qquad (16)$$

where $V_p$ is the volume of the particle. It has been shown in a series of papers by the group of S. Arnold and the others that the polarizability of a rod-like particle (i.e., similar to hemozoin crystals in this work) is different for different polarizations of the resonator field as well as for different orientations of the particle on the resonator surface (i.e., when it is standing or lying on the surface for fixed polarization) [33-37]. For a rod-like particle making an angle $\theta$ with the surface of the resonator and $\varphi$ around its axis, polarizabilities are given as

$$\alpha_{TE} = \xi_1 \left(1 - \sin^2\theta \cos^2\varphi\right) + \xi_2 \sin^2\theta \cos^2\varphi \qquad (17)$$

$$\alpha_{TM} = \xi_2 \sin^2\theta + \xi_2 \cos^2\theta \qquad (18)$$

where $\xi_1 = 2(n_p^2 - n_e^2)/(n_p^2 + n_e^2)$ and $\xi_2 = 1 - n_e^2/n_p^2$. Rod-like particle is lying on the surface when $\theta=0$, and standing on the surface when $\theta=\pi/2$. The latter is highly unlikely to be observed when a rod-like particle is deposited on microtoroid resonator. Thus, the observed mode splitting spectrum in our experiments should also change with the cavity mode and the orientation of the particles/crystals because of its dependence on the polarizability via the defining expressions $2|g|=\alpha f^2(r)\omega/V$ and $2\Gamma=2\alpha|g|\omega^3/3\pi v^3$. However, in our experiments we could not see any significant change in the estimated polarizabilities when the polarization was changed. In experiments, we observe that TE and TM modes of a resonator have slightly different resonance wavelengths. Moreover, numerical simulations with COMSOL reveals that they also have different field distributions as well as slightly different mode volumes; thus $2|g|$ and $2\Gamma$ will be affected by the polarization of the mode and the orientation of the particle. However, when we take their ratio as we did for size estimation, arriving at Eq. (15), the dependence on generally unknown parameters $f^2(r)$ and $V$ is eliminated. For a particle deposited on the resonator with a fixed orientation, experiments with TE and TM modes will reveal $\alpha_t=\alpha_{TE}$ and $\alpha_n=\alpha_{TM}$ from the experimentally observed spectra and known resonance wavelengths of the TE and TM modes. Denoting the experimentally observed splitting amount and the linewidth differences as $2|g_{TE}|$ and $2\Gamma_{TE}$ for the TE mode and $2|g_{TM}|$ and $2\Gamma_{TM}$ for the TM mode, we arrive at

$$\frac{\alpha_{TE}}{\alpha_{TM}} = \left(\frac{\lambda_{TE}}{\lambda_{TM}}\right)^3 \frac{\Gamma_{TE}}{\Gamma_{TM}} \frac{|g_{TM}|}{|g_{TE}|} \sim \frac{\Gamma_{TE}}{\Gamma_{TM}} \frac{|g_{TM}|}{|g_{TE}|} \qquad (19)$$

where we have used the observation that the resonance wavelengths $\lambda_{TE}$ and $\lambda_{TM}$ of the TE and TM modes differ only slightly and hence their ratio can be approximated to *1*. Therefore, we should be able to resolve any change in the splitting spectrum when the mode is changed from TE to TM or vice versa if the particle has a non-spherical shape and the polarizabilities for TE and TM modes are sufficiently large. In Fig. 8, we show $\alpha_{TE}/\alpha_{TM}$ for a spherical particle as well as for a rod-like particle with various orientations $\theta$ and $\varphi$ as a function of $n_p$ the refractive index of the particle assuming that the volume $V_p$ of the particle is fixed. It is clear that $\alpha_{TE}/\alpha_{TM}$ is always *1* for a spherical particle regardless of $n_p$ its refractive index. However, for a rod-like particle, which is depicted as a nano-cylinder here, $\alpha_{TE}/\alpha_{TM}$ changes with the orientation and refractive index of the particle. It is interesting to see that as the particle refractive index decreases, $\alpha_{TE}/\alpha_{TM}$ decreases for all orientations of the particle. For a rod-like particle having refractive index *1.44* (i.e., this is the refractive index for the hemozoin crystals) or smaller, $\alpha_{TE}/\alpha_{TM}$ is very close to *1* suggesting that for such rod-like particles, it is very difficult to discriminate among orientation dependent polarizability values. For example for $\varphi=0$, we find $\alpha_{TE}/\alpha_{TM}$ equals to *1.079* and *0.93*, for $\theta=0$ (lying on the surface) and $\theta=\pi/2$ (standing on the surface), respectively. Similarly, for $\varphi=\pi/2$, we find $\alpha_{TE}/\alpha_{TM}$ equals to *1.079* and *1*, for $\theta=0$ and $\theta=\pi/2$, respectively. It is clear that for the non-extreme orientations, $\alpha_{TE}/\alpha_{TM}$ approaches to *1* and it becomes much more difficult to measure or observe the changes. Detecting such small differences of the resonance frequencies and linewidths of split modes with significant amount of noise (e.g., electrical noise, laser noise, and the signal processing noise) is very difficult. We believe that this is the main reason why we could not observe change in the estimated polarizabilities.

In an experiment, $\alpha_{TE}/\alpha_{TM}$ is estimated by comparing the mode splitting spectra obtained for TE and TM modes. In order to do this estimation, the first condition is to be able to resolve mode splitting spectra for the same particle when the mode is TE and TM according to the resolvability criterion $2|g|>\Gamma+\omega/Q$. We performed numerical simulations using COMSOL and obtained resonance wavelengths, quality factors, mode volumes, and field distributions for a microtoroid of similar size that we used in the experiments as a function of the volume of a rod-like particle with refractive index of *1.44* and $\varphi=0$. For each particle volume, we estimated $\alpha_{TE}$ and $\alpha_{TM}$ from Eqs. 17 and 18, and then used these values and parameters estimated from COMSOL simulations in Eqs. 12 and 13 to find $2|g_{TE}|$ and $\Gamma_{TE}$ for the TE mode and $2|g_{TM}|$ and $\Gamma_{TM}$ for the TM mode and plotted them in Fig. 9. Note that in these simulations, we chose and the location of the particle in the field such that maximal values are obtained for the depicted parameters. Even under these optimal conditions, it is clear that when the volume of the rod-like particle is small, it becomes very difficult to observe differences in these measurable parameters. In an experiment, additional noises will make it even more difficult to observe small changes in these measurable parameters. In short we believe that due to the low refractive index contrast between the hemozoin crystal and the water, and the small size (volume) of the hemozoin crystals the polarizability values for TE and TM modes are already very small. Moreover, the possible orientations that these crystals take on the resonator surface do not maximize polarization dependent poalrizability changes so that such changes are observed in noisy experimental conditions. It may be the case that if we deposit sufficiently large number of crystals to achieve a higher resonator surface coverage, then such small changes may be amplified to be observed.

## 4. CONCLUSIONS

In conclusion, we report the first single nanoparticle detection and size measurement in aquatic environment using mode splitting in a WGM resonator. We have achieved detecting and measuring PS nanoparticles of nominal radii *75nm*. In addition, we tested the performance of our platform using synthetic hemozoin crystals, showing the detection and size measurement of individual hemozoin crystals. Moreover, we have characterized splitting quality factor that can be used to determine mode splitting observability in response to light scatterers, such as nanoparticles, in the mode volume of a resonator and obtained the relation between the quality factor of a resonance mode and the noise level in the system, which have been used to identify operating regimes of a WGM resonator: (i) The non-degenerate region where mode splitting takes place, lifting the degeneracy and leading to a doublet in the transmission spectra upon binding of a particle, and (ii) The degenerate region where transmission spectra shows a single resonance which undergoes reactive shift or linewidth broadening upon the binding of a particle to the resonator. Within each of these regions, there are sub-domains within which noise level in the system determines whether accurate size measurement can be done or not. We have identified the linewidth measurement noise as the main source of error in the accuracy of size measurement.

We have found that resonators with high-$Q$ modes favor mode splitting (non-degenerate regime) whereas those with lower $Q$ favor reactive shift and single mode operation (degenerate regime). We believe that the findings of this paper provide an answer to the ongoing debate on mode splitting and spectral shift methods, and can be used as a guide for the future studies.

Control and desired eradication of malaria parasites requires rapid and sensitive detection techniques for immediate and effective treatment. This study for the first time shows that detection of single hemozoin crystals, which are the signature of malaria parasites, can be achieved by WGM. Therefore, detection and measurement of individual hemozoin crystals in aquatic environment opens the way for highly sensitive early-phase detection of malaria infection from serum.

These results bring us one step closer to single particle or single molecule detection in aquatic environment using mode splitting. The next task is to functionalize the resonator surface to selectively detect hemozoin crystals or other biological nanoparticles in serum or blood.

**Acknowledgements**

We would like to thank Dr. T. Tsukui and Y. Igari from Zenoaq (ZENOAQ, Nippon Zenyaku Kogyo Co. Ltd., Koriyama, Fukushima, Japan) for their help for hemozoin preparation, K. Ohata for technical assistance for taking SEM pictures. The authors acknowledge supports from the National Science Foundation (Grant No. 0954941), the Japanese Ministry of Education, Culture, Sports, Science and Technology (MEXT) and Japan Science and Technology Agency (JST). This work is performed in part at the NRF/NNIN (NSF award No. ECS-0335765) of Washington University in St. Louis.

**REFERENCES**

1. M. Li, H. X. Tang, and M. L. Roukes, "Ultra-sensitive NEMS-based cantilevers for sensing, scanned probe and very high-frequency applications," Nat Nanotechnol.**2**(2), 114-20 (2007).
2. J. Homola, "Surface plasmon resonance sensors for detection of chemical and biological species," Chem. Rev. **108**(2), 462-93 (2008).
3. F. Vollmer, D. Braun, A. Libchaber, M. Khoshsima, I. Teraoka, and S. Arnold, "Protein detection by optical shift of a resonant microcavity," Appl. Phys. Lett. **80**(21), 4057-4059 (2002).
4. P. F. Scholl, D. Kongkasuriyachai, P. A. Demirev, A. B. Feldman, J. S. Lin, D. J. Sullivan, Jr., and N. Kumar, "Rapid detection of malaria infection in vivo by laser desorption mass spectrometry," Am J Trop Med Hyg. **71**(5), 546-51 (2004).
5. T. Hanscheid, T. J. Egan, and M.P. Grobusch, "Haemozoin: from melatonin pigment to drug target, diagnostic tool, and immune modulator," Lancet Infect Dis. **7**(10), 675-85 (2007).
6. C. Coban, M. Yagi, K. Ohata, Y. Igari, T. Tsukui, T. Horii, K. J. Ishii, and S. Akira, "The malarial metabolite hemozoin and its potential use as a vaccine adjuvant," Allergol Int**. 59**(2), 115-24 (2010).
7. J. M. Belisle, S. Costantino, M. L. Leimanis, M. J. Bellemare, D. S. Bohle, E. Georges, and P. W. Wiseman, "Sensitive detection of malaria infection by third harmonic generation imaging," Biophys J. **94**(4), L26-8 (2008).
8. B. R. Wood, E. Bailo, M. A. Khiavi, L. Tilley, S. Deed, T. Deckert-Gaudig, D. McNaughton, and V. Deckert, "Tip-enhanced Raman scattering (TERS) from hemozoin crystals within a sectioned erythrocyte," Nano Lett. **11**(5), 1868-73 (2011).
9. F. Vollmer, S. Arnold, D. Braun, I. Teraoka, and A. Libchaber, "Multiplexed DNA quantification by spectroscopic shift of two microsphere cavities," Biophys J. **85**(3), 1974-1979 (2003).
10. F. Vollmer, S. Arnold, and D. Keng, "Single virus detection from the reactive shift of a whispering-gallery mode," P. Natl. Acad. Sci. USA **105**(52), 20701-20704 (2008).
11. F. Vollmer and S. Arnold, "Whispering-gallery-mode biosensing: label-free detection down to single molecules," Nat Methods **5**(7), 591-596 (2008).
12. S. Arnold, D. Keng, S. I. Shopova, S. Holler, W. Zurawsky, and F. Vollmer, "Whispering gallery mode carousel-a photonic mechanism for enhanced nanoparticle detection in biosensing," Opt. Express **17**(8), 6230-8 (2009).
13. J. Zhu, S. K. Ozdemir, Y. F. Xiao, L. Li, L. He, D. Chen, and L. Yang, "On-chip single nanoparticle detection and sizing by mode splitting in an ultrahigh-Q microresonator," Nat Photonics **4**(1), 46-49 (2010).
14. W. Kim, S. K. Ozdemir, J. Zhu, L. He, and L. Yang, "Demonstration of mode splitting in an optical microcavity in aqueous environment," Appl. Phys. Lett. **97**(7), 071111 (2010).
15. W. Kim, S. K. Ozdemir, J. Zhu, and L. Yang, "Observation and characterization of mode splitting in microsphere resonators in aquatic environment," Appl. Phys. Lett. **98**(14), 141106 (2011).
16. T. Lu, H. Lee, T. Chen, S. Herchak, J. H. Kim, S. E. Fraser, R. C. Flagan, and K. Vahala, "High sensitivity nanoparticle detection using optical microcavities," P. Natl. Acad. Sci. USA**108**(15), 5976-5979 (2011).
17. L. He, K. Ozdemir, J. Zhu, W. Kim, and L. Yang, "Detecting single viruses and nanoparticles using whispering gallery microlasers," Nat Nanotechnol **6**(7), 428-432 (2011).
18. D. K. Armani, T. J. Kippenberg, S. M. Spillane, and K. J. Vahala, "Ultra-high-Q toroid microcavity on a chip," Nature **421**(6926), 925-928 (2003).

19. S. Pagola, P. W. Stephens, D. S. Bohle, A. D. Kosar, and S. K. Madsen, "The structure of malaria pigment beta-haematin," Nature **404**(6775), 307-10 (2000).
20. C. Coban, Y. Igari, M. Yagi, T. Reimer, S. Koyama, T. Aoshi, K. Ohata, T. Tsukui, F. Takeshita, K. Sakurai, T. Ikegami, A. Nakagawa, T. Horii, G. Nunez, K. J. Ishii, and S. Akira, "Immunogenicity of whole-parasite vaccines against Plasmodium falciparum in volves malarial hemozoin and host TLR9," Cell Host Microbe **7**(1), 50-61 (2010).
21. B. Y. H. Liu and D. Y. H. Pui, "A submicron aerosol standard and the primary, absolute calibration of the condensation nucleicounter," J. Colloid Interface Sci. **47**(1), 155-171 (1974).
22. E. O. Knutson and K. T. Whitby, "Aerosol classification by electric mobility: apparatus, theory, and applications," J. Aerosol. Sci. **6**(6), 443-451 (1975).
23. J. Schurr and K. Schmitz, "Dynamic light scattering studies of biopolymers: effects of charge, shape, and flexibility," Ann. Rev. Phys. Chem. **37**(271), (1986).
24. William C. Hinds, *Aerosol Technology: Properties, Behavior, and Measurement of Airborne Particles* (John Wiley & Sons, 2nd ed., New York, 1999).
25. A. Mazzei, S. Goetzinger, L. D. Menezes, G. Zumofen, O. Benson, and V. Sandoghdar, "Controlled coupling of counter propagating whispering-gallery modes by a single Rayleigh scatterer: A classical problem in a quantum optical light," Phys. Rev. Lett. **99**(17), 173603 (2007).
26. M. L. Gorodetsky, A. D. Pryamikov, and V. S. Ilchenko, "Rayleigh scattering in high-Q microspheres," J. Opt. Soc. Am. B **17**(6), 1051-1057 (2000).
27. V. S. Ilchenko and M. L. Gorodetsky, "Thermal nonlinear effects in optical whispering gallery microresonators," Laser Phys. **2**(6), 1004-1009 (1992).
28. J. Zhu, S. K. Ozdemir, L. He, D. Chen, and L. Yang, "Single virus and nanoparticle size spectrometry by whispering-gallery-mode microcavities," Opt. Express 19(17), 16195-16206 (2011).
29. Jiangang Zhu, *Ultra-high-Q microresonator with applications towards single nanoparticle sensing* (Washington university in St. Louis, Ph.D dissertation, 2011).
30. S. I. Shopova, R. Rajmangal, Y. Nishida, and S. Arnold, "Ultrasensitive nanoparticle detection using a portable whispering gallery mode biosensor driven by a periodically poled lithium-niobate frequency doubled distributed feedback laser," Rev. Sci. Instrum. **81**(10), 103110 (2010).
31. S. K. Ozdemir, J. Zhu, L. He, and L. Yang, "Estimation of Purcell factor from mode-splitting spectra in an optical microcavity," Phys. Rev. A **83**(3), 033817 (2011).
32. Y. Park, M. Diez-Silva, G. Popescu, G. Lykotrafitis, W. Choi, M. S. Feld, and S. Suresh, "Refractive index maps and membrane dynamics of human red blood cells parasitized by Plasmodium falciparum," Proc. Natl. Acad. Sci. USA **105**(37), 13730-5 (2008).
33. S. Arnold, M. Khoshsima, I. Teraoka, S. Holler, and F. Vollmer, "Shift of whispering-gallery modes in microspheres by protein adsorption," Opt. Lett. **28**(4), 272-274 (2003).
34. I. Teraoka and S. Arnold, "Theory of resonance shifts in TE and TM whispering gallery modes by nonradial perturbations for sensing applications," J. Opt. Soc. Am. B **23**(7), 1381-1389 (2006).
35. I. Teraoka, S. Arnold, and F. Vollmer, "Perturbation approach to resonance shifts of whispering-gallery modes in a dielectric microsphere as a probe of a surrounding medium," J. Opt. Soc. Am. B **20**(9), 1937-1946 (2003).
36. M. Noto, D. Keng, I. Teraoka, and S. Arnold, "Detection of protein orientation on the silica microsphere surface using transverse electric/transverse magnetic whispering gallery modes," Biophys J. **92**(12), 4466-4472 (2007).
37. X. Yi, Y. F. Xiao, Y. Li, Y. C. Liu, B. B. Li, Z. P. Liu, and Q. H. Gong, "Polarization-dependent detection of cylinder nanoparticles with mode splitting in a high-Q whispering-gallery microresonator," Appl. Phys. Lett. **97**(20), (2010).

# FIGURES

**FIGURE 1:**

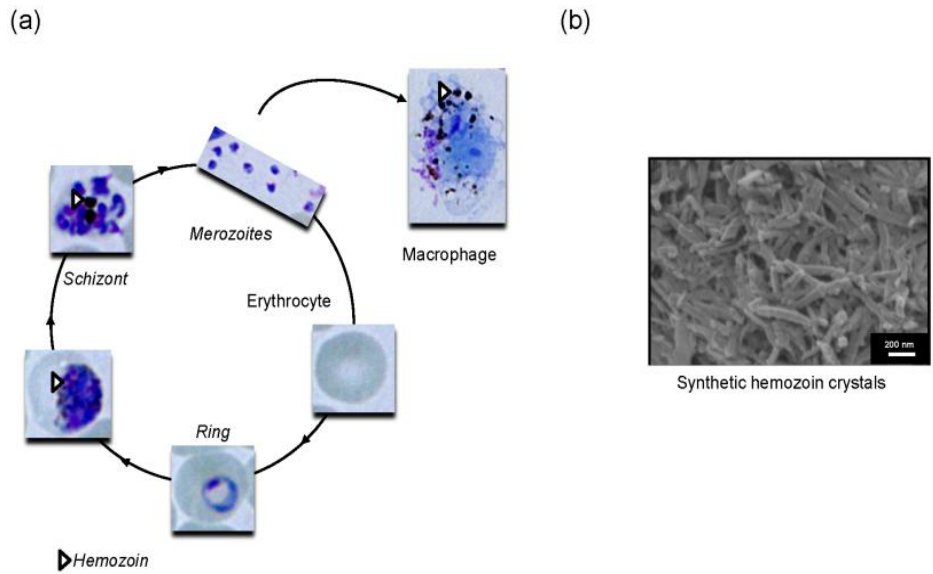

**Fig. 1.** (a) Hemozoin is continuously produced by *Plasmodium* parasites inside the erythrocytes. *Plasmodium* parasites have a repetitive life cycle within the erythrocytes of the mammalian host (erythrocytic blood-stage). Hemozoin is produced primarily within trophozoite-stage parasites and is almost continuously released into the blood stream during malaria infection. Released hemozoin is readily captured by immune system cells such as macrophages. Figure shows the life cycle of rodent parasite *P. yoelii*. (b) SEM pictures of synthetic hemozoin purified from hemin chloride.

**FIGURE 2:**

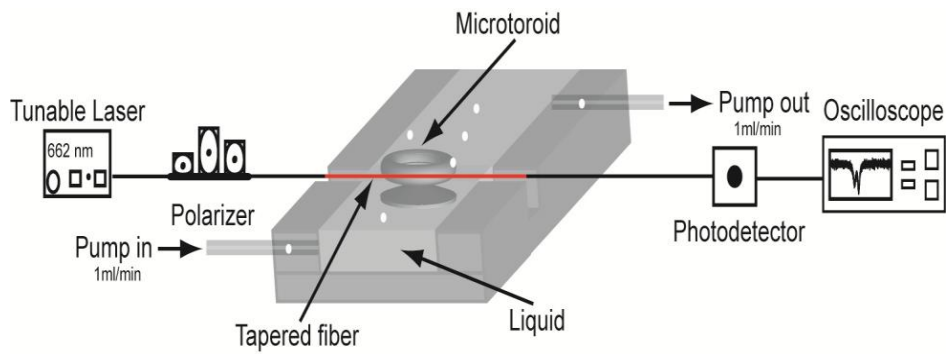

**Fig. 2.** Schematics of the setup used in mode-splitting based single particle detection and measurement experiments in aquatic environment. A high-*Q* whispering gallery mode microtoroid resonator is placed in an aquatic chamber (dimension: *3.2cm×1cm×0.5cm*). The solution including hemozoin crystals or nanoparticles is pumped into and out of the chamber at a rate of *1ml/min* using a micropump.

**FIGURE 3:**

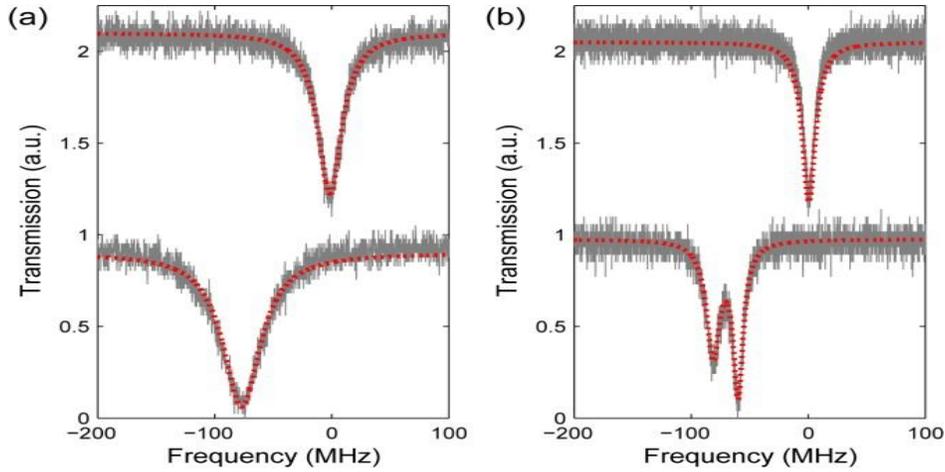

**Fig. 3.** Reactive shift (resonance shift) and mode splitting spectra. As a response to a binding particle, a single WGM resonance (a) shifts to a lower frequency and experiences linewidth broadening or (b) undergoes mode splitting where two spectrally shifted resonances of different linewidths are obtained in the transmission spectra. Upper spectra are obtained before particle binding.

**FIGURE 4:**

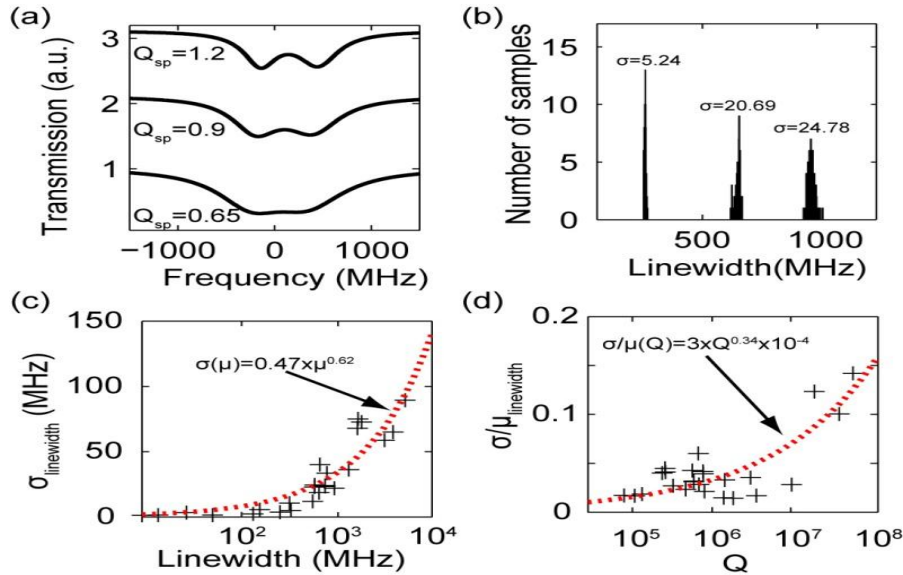

**Fig. 4.** (a) Resolvability of the split modes in the transmission spectra and its relation with the splitting quality, $Q_{sp}$. The higher the $Q_{sp}$, the better the resolvability of the doublets. (b) Standard deviation, $\sigma$, of the linewidth measurements obtained for three WGM resonances of different quality factors (linewidths) in the same microtoroid resonator in water. Standard deviation is higher for resonances of lower quality factors (larger linewidths). (c) Dependence of the standard deviations of the estimated linewidths as a function of the average linewidth $\mu$. (d) Relation between the quality factor of a WGM resonance and coefficient of variation $\sigma/\mu$.

**FIGURE 5:**

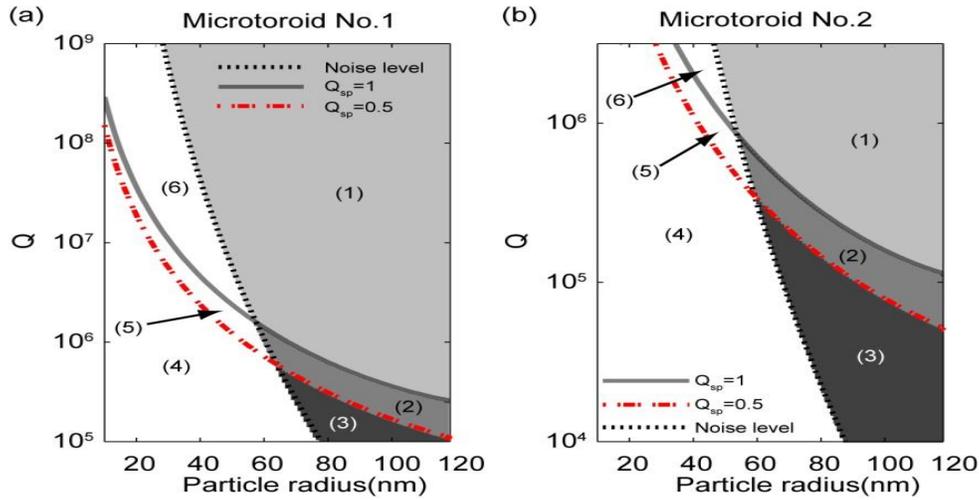

**Fig. 5.** A WGM resonator will experience either reactive shift or mode splitting depending on the size of the binding PS particle, quality factor of the resonance, and the noise level in the system. There are four possible regions: (1) mode splitting with highly accurate size measurement, (2) mode splitting but with erroneous size measurement, (3) reactive shift (mode splitting cannot be resolved or does not take place) with accurate size measurement, and (4) reactive shift with erroneous size measurement. The areas of these regions depend on the diameter (*D*) of the resonator, (a) *D=80μm*, and (b) *D=53μm*. Dotted line shows the noise level measured in our experiments.

**FIGURE 6:**

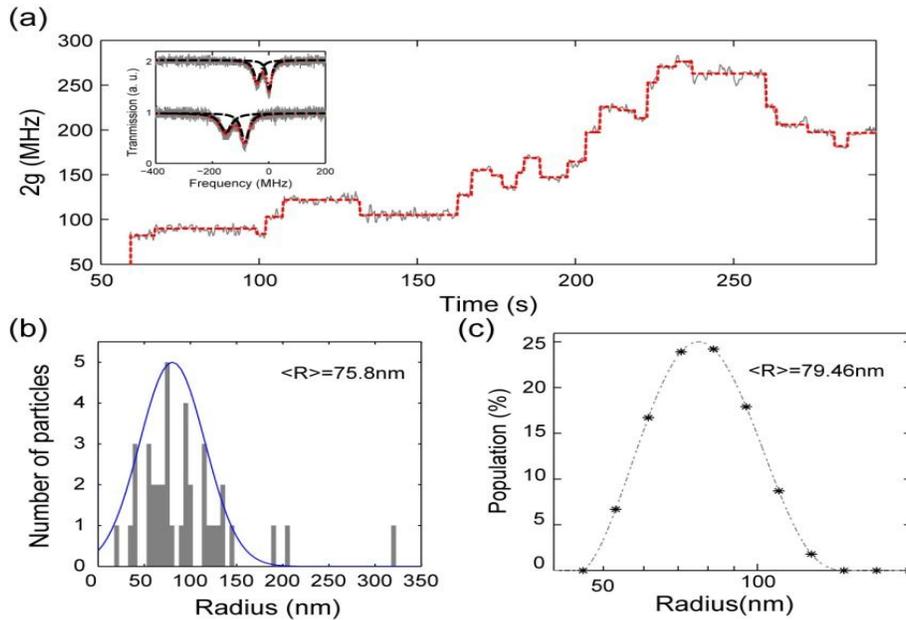

**Fig. 6.** Mode splitting based detection and single-shot size measurement of individual polystyrene (PS) particles of nominal size *R=75±2.2nm* binding to a microtoroid placed in aquatic environment. (a) Change in the amount of mode splitting as the particles enter the mode volume of a microtoroid one by one. Inset depicts typical particle induced transitions from single resonance to doublets (split resonances) in the transmission spectra. (b) Distribution of the estimated size of detected particles. (c) Size distribution obtained from dynamic light scattering measurements (DLS) for the same particles.

**FIGURE 7:**

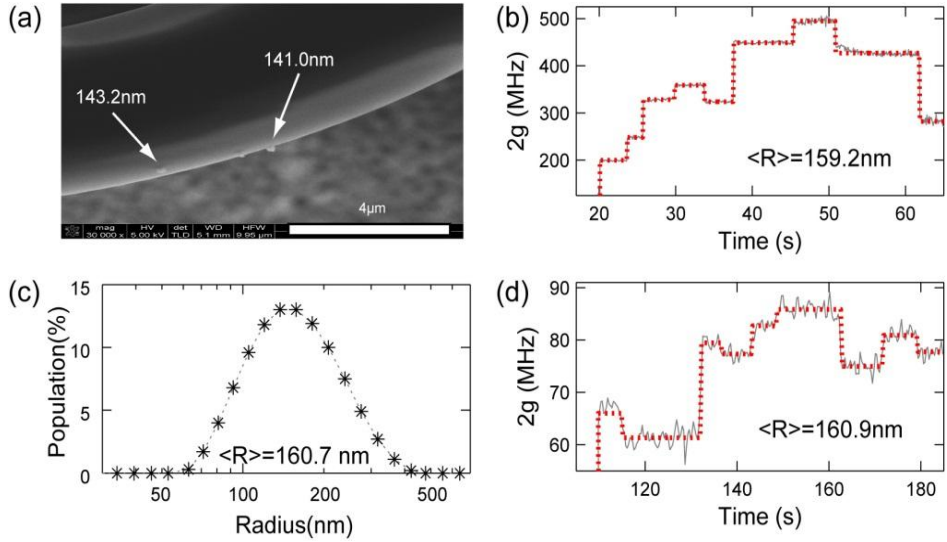

**Fig. 7.** Detection and measurement of hemozoin crystals. (a) SEM image of hemozoin crystals deposited on a microtoroid resonator. (b) The amount of mode splitting induced by consecutively deposited hemozoin crystals on a microtoroid. Experiments were done in air. (c) Typical size distribution obtained from DLS measurements for the hemozoin crystals. (d) Result of mode splitting experiment performed in aquatic environment for hemozoin crystals. Discrete jumps in (b) and (d) signals that a hemozoin is within the mode volume of the resonator.

**FIGURE 8:**

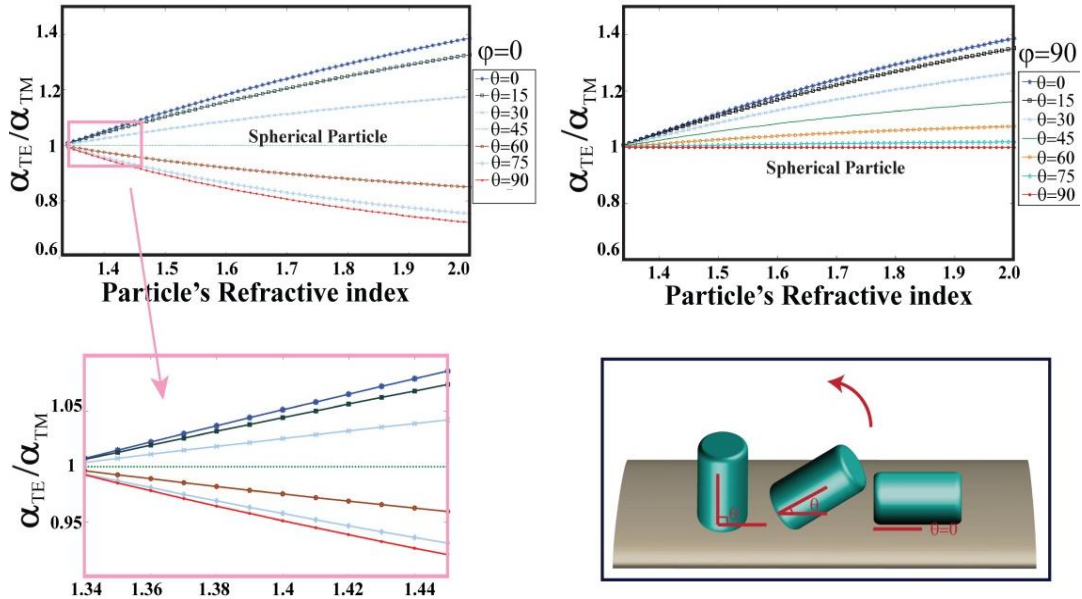

**Fig. 8.** The ratio $\alpha_{TE}/\alpha_{TM}$ of the polarizabilities $\alpha_{TE}$ and $\alpha_{TM}$ of a rod-like particle with fixed volume making an angle $\theta$ with the resonator surface and orientation $\varphi$ when the field is TE and TM, respectively, as a function of the refractive index of the particle.

**FIGURE 9:**

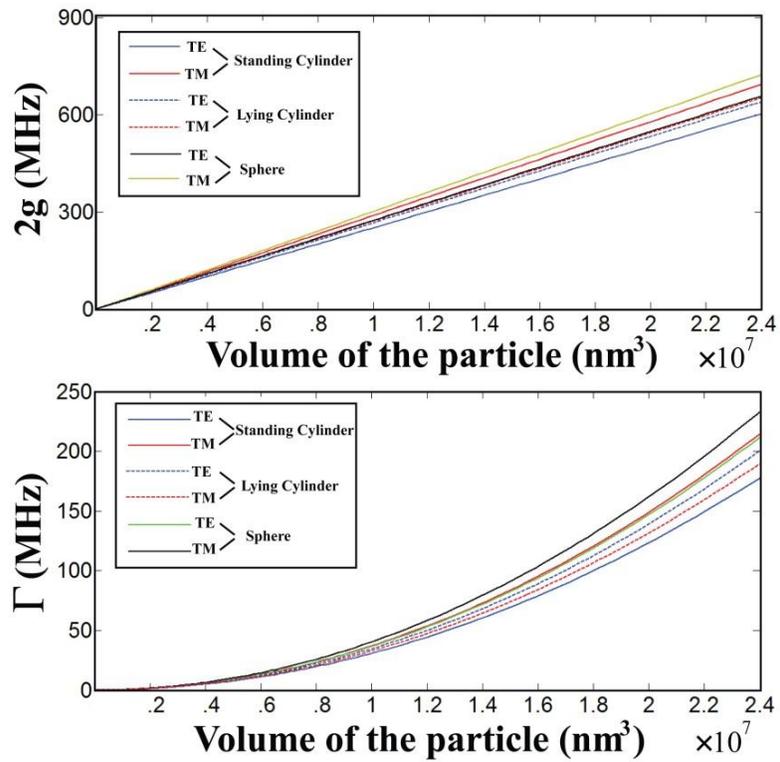

**Fig. 9.** Numerical simulation results showing the dependence of the amounts of mode splitting and the linewidth broadening induced by a single rod-like particle as a function of its volume for different polarizations of the resonator field and the orientation of the particle on the resonator.